\documentclass[aps,prl,twocolumn,superscriptaddress]{revtex4-2}

\usepackage{graphicx}
\usepackage{amsmath}
\usepackage{amssymb}
\usepackage{bm}
\usepackage{xcolor}
\usepackage{hyperref}
\usepackage{booktabs}   
\usepackage{multirow}   
\hypersetup{
  colorlinks=true,
  breaklinks=true,
  linkcolor=blue,
  urlcolor=blue,
  citecolor=blue,
}
\usepackage{changes}
\usepackage{comment}
\usepackage{ulem}
\def\<#1>{\mathinner{\langle#1\rangle}}
\mathchardef\up="0222
\mathchardef\dn="0223

\newcommand{\kh}[1]{{\color{black}#1}}

\newcommand{\khx}[1]{}

\newcommand{\ej}[1]{{\color{black}#1}}

\newcommand{\lv}[1]{{\color{black}#1}}

\begin{document}
\preprint{arXiv}

\title{\kh{Beyond the conventional Emery model: \\ crucial role of long-range hopping for cuprate superconductivity
}}

\author{Eric Jacob}
\email{eric.jacob@tuwien.ac.at}
\affiliation{Institute of Solid State Physics, TU Wien, 1040 Vienna, Austria}

\author{M.~O.~Malcolms}
\affiliation{Max-Planck-Institut für Festkörperforschung, Heisenbergstraße 1, 70569 Stuttgart, Germany}

\author{Viktor Christiansson}
\affiliation{Institute of Solid State Physics, TU Wien, 1040 Vienna, Austria}

\author{Leonard M. Verhoff}
\affiliation{Institute of Solid State Physics, TU Wien, 1040 Vienna, Austria}

\author{Paul Worm}
\affiliation{Institute of Solid State Physics, TU Wien, 1040 Vienna, Austria}

\author{Liang Si}
\affiliation{School of Physics, Northwest University, Xi'an 710127, China}
\affiliation{Institute of Solid State Physics, TU Wien, 1040 Vienna, Austria}

\author{Philipp Hansmann}
\affiliation{Department of Physics, Friedrich-Alexander-Universit\"at Erlangen/N\"urnberg, 91058 Erlangen, Germany}
\affiliation{Science Institute and Faculty of Physical Sciences, University of Iceland, 107 Reykjav\'ik, Iceland}

\author{Thomas~Sch{\"a}fer}
\affiliation{Max-Planck-Institut für Festkörperforschung, Heisenbergstraße 1, 70569 Stuttgart, Germany}
\affiliation{Dipartimento di Fisica, Università di Trieste, Strada Costiera 11, I-34151 Trieste, Italy}

\author{Karsten Held}
\email{held@ifp.tuwien.ac.at}
\affiliation{Institute of Solid State Physics, TU Wien, 1040 Vienna, Austria}

\begin{abstract}
The  Emery model is the quintessential model
for cuprate superconductors.
In his eponymous paper, Emery only considered
the next-nearest-neighbor oxygen-copper hopping.
Later, also the relevance of nearest- and next-nearest
oxygen-oxygen hoppings has been pointed out. Using
dynamical vertex approximation\khx{for calculating the superconducting phase diagram of the Emery model}, 
we find a \kh{superconducting dome consistent with cuprates. However,} long-range hoppings beyond the \kh{three} conventional hopping parameters are necessary \kh{for the quantitatively correct phase diagram and for a proper $d$-wave order parameter.} 
\end{abstract}

\maketitle

\begin{figure}[tb!]
\centering
\includegraphics[width=1.\columnwidth]{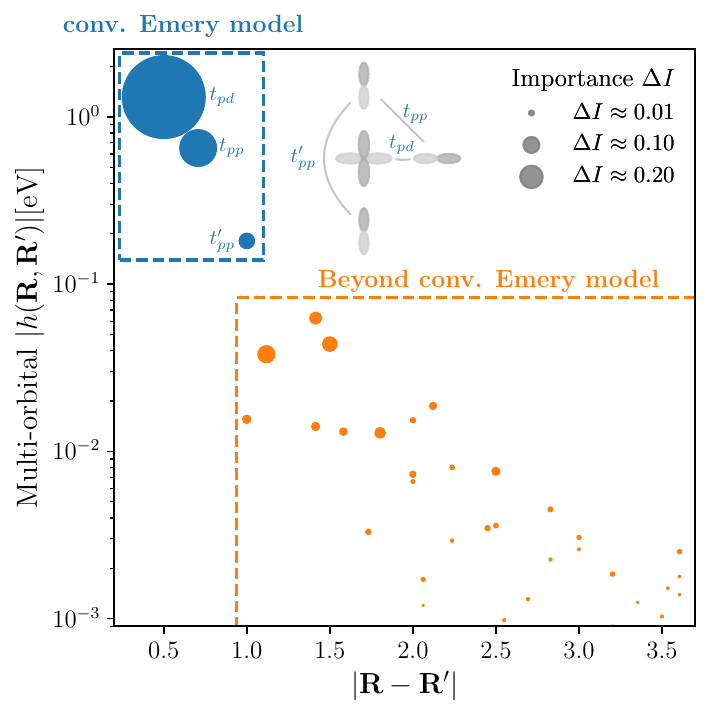}
\vspace{-.5cm}
\caption{Hopping elements of the three-band Wannier Hamiltonian for CaCuO$_2$ and their importance $\Delta I$ 
({dot area; definition see text}) as a function of distance ${\mathbf R}-{\mathbf R}'$. 
The three blue hoppings of the sketch and within the dashed blue box are those hitherto considered in the conventional Emery model.}
\label{fig:long_range_hoppings}
\end{figure}

\textit{Introduction.} 
Cuprates are charge-transfer insulators according to
the scheme of Zaanen, Sawatzky and Allen~\cite{Zaanen1985}. That is, doped holes primarily
go into the oxygen orbitals, as experimentally confirmed \ej{\cite{gauquelin_2014,Jurkutat2014,Chen_1991}}. This makes the Emery model \cite{Emery1987} with the copper $d_{x^2-y^2}$ and two in-plane oxygen orbitals (see inset of Fig.~\ref{fig:long_range_hoppings}) 
the
{quintessential}  model for superconducting cuprates \cite{Bednorz1986}.

{Despite this multi-orbital nature,} oxygen and copper orbitals lead to a single
Fermi surface \cite{Damascelli2003} of mixed copper-oxygen character. This motivates the use of the 
one-band Hubbard model instead of the more involved three-band Emery model. Indeed, most theoretical studies of cuprate superconductivity are based on this Hubbard model \cite{Gull2015,Qin21,Arovas2022}. On a qualitative level, such a simplified calculation with a Hubbard model is reasonable;
however, quantitatively reliable calculations
with a single set of parameters across dopings require the more fundamental Emery model \cite{Tseng2025}.

The Emery model has been studied with various methods, including -- most closely to our study --
dynamical mean-field theory \cite{Zolfl2000,Weber2008,Hansmann2014,Han2021,Tseng2025,Vucicevic2026},
its cluster \cite{Weber2012,Fratino2016b,Kowalski2021,Mai2021,Mai2021b,Sordi2024,Reaney2025,Kumar2025} and diagrammatic extensions \cite{Malcolms2026}{, as well as the two-particle self-consistent (TPSC) approach~\cite{TPSC_Emery_2024}.}

{Cluster extensions of DMFT \cite{Weber2012,Fratino2016b,Kowalski2021,Mai2021,Mai2021b,Sordi2024,Reaney2025,Kumar2025,Bacqlabreuil2025} and direct lattice quantum Monte Carlo simulations \cite{Huang2017Stripes, Mai2024Fluctuating,Peng2025} are, in principle, numerically
exact, when extrapolated to infinite cluster size.
However, the sign problem renders this extrapolation infeasible. Typically, 
only 2$\times$2 clusters have been used for low temperature simulations of the Emery model around $T_c$ \cite{Mai2021,Kowalski2021}, and are known to strongly overestimate superconducting tendencies. Larger clusters such as 
6$\times$6 \cite{Mai2021}, 8$\times$8 \cite{Peng2025} and 16$\times$4 \cite{Huang2017Stripes, Mai2024Fluctuating} have been studied \kh{-- but only} at elevated temperatures. Even these clusters are definitely too small to \kh{properly resolve Fermi surfaces, which is crucial for an accurate description of superconductivity that gaps them. 
Additionally, some small cluster geometries have the tendency to introduce spurious modulations due to their rectangular shapes.}

{In contrast, diagrammatic extensions of DMFT~\cite{RMPVertex} 
allow for fine momentum grids of ${\cal O}(10^6)$  $\mathbf k$-points, essentially corresponding to an infinitely large system.  
These diagrammatic extensions and, in particular, the dynamical vertex approximation (D$\Gamma$A \cite{Toschi2007,Katanin2009}) have been employed to study superconductivity in the Hubbard model \cite{Kitatani2019} and infinite-layer nickelates  \cite{Kitatani2020}, even successfully predicting the
experimental phase diagram~\cite{Lee2023}.  
The approximation underlying the diagrammatic extensions relies on the momentum-independence of the {\em irreducible} vertex.
This assumption is justified over a wide parameter range; however, at very small dopings 
the (three point) vertex \kh{also acquires a significant fermionic momentum dependence, i.e., is much larger at the antinode \cite{Yu2025}.}

}

In this Letter, we employ D$\Gamma$A to study \kh{the Emery model and obtain a superconducting phase diagram akin to cuprates}.
We find that the hopping terms {conventionally} used in virtually all Emery calculations so far (blue box in Fig.~\ref{fig:long_range_hoppings}) are \emph{insufficient} for $d$-wave superconductivity \kh{beyond 15\% hole doping}. 
Long-range hopping parameters (orange box in Fig.~\ref{fig:long_range_hoppings}) are needed for a \kh{correct} description. {Some of t}hese hopping elements are actually of the same range as the next-nearest-neighbor hopping $t'$ in the Hubbard model, which is known to be important for superconductivity \cite{Pavarini2001,Qin2020}.
\kh{Without these long-range hopping parameters we also obtain a strongly modulated $d$-wave superconducting order
parameter against conventional wisdom for cuprates.}

\textit{Model and Method.}
We study various two-dimensional $d$–$p$ models defined by the Hamiltonian
\begin{equation}
H = \sum_{\mathbf{k},\sigma} 
\psi^{\dagger}_{\mathbf{k},\sigma}\,\bar{h}_{0}(\mathbf{k})\,\psi_{\mathbf{k},\sigma}
+ U\sum_{i} n^{d}_{i,\uparrow} n^{d}_{i,\downarrow},
\label{eq:hamiltonian}
\end{equation}
where 
$\psi^{\dagger}_{\mathbf{k},\sigma}
=(d^{\dagger}_{\mathbf{k},\sigma},
p^{\dagger}_{x,\mathbf{k},\sigma},
p^{\dagger}_{y,\mathbf{k},\sigma})$
contains the electronic creation operators for the copper
$d_{x^{2}-y^{2}}$ orbital and the oxygen $p_x$ and $p_y$ orbitals
with spin $\sigma$ and momentum $\mathbf{k}=(k_x,k_y)$;  {$\psi_{\mathbf{k},\sigma}$ is the corresponding annihilation operator.}
The momentum-dependent $3\times3$ matrix $\bar{h}_{0}(\mathbf{k})$
defines the noninteracting band dispersion, and $U$ denotes the local
on-site Coulomb repulsion on the copper $d$ orbital{; $n^{d}_{i,\sigma}= d^{\dagger}_{i,\sigma} d_{i,\sigma}$  is the occupation number operator for spin $\sigma$ on site $i$.}

To obtain $\bar{h}_{0}(\mathbf{k})$ we first \lv{perform} density-functional theory (DFT) calculations for \kh{the arguably simplest cuprate superconductor, CaCuO$_2$,} 
with  $P4/mmm$ space group (No.~123) and lattice parameters $a=b=3.85$\,{\AA}, \lv{and} $c=3.14$\,{\AA}. More specifically, we use the generalized gradient approximation (GGA) with Perdew-Burke-Ernzerhof (PBE) exchange-correlation functional \cite{Perdew96} as implemented in the \textsc{WIEN2k} code \cite{blaha2001wien2k}  and a $19\!\times\!19\!\times\!23$ $\mathbf k$-mesh.
In a second step, the DFT band structure is projected onto maximally localized Wannier functions \cite{RevModPhys.84.1419}, employing the
\textsc{wien2wannier} \cite{Kunes2010a} and \textsc{Wannier90} \cite{Pizzi2020} codes. 
.

The {DFT-Wannier} $\bar{h}_{0}(\mathbf{k})$ is supplemented by an interaction  $U = 6.5\,\mathrm{eV}$, which we calculated using the constrained random phase approximation (cRPA) \cite{Aryasetiawan2004}, removing the three bands of the $d$-$p$ model from the screening processes. We use the static value of $U$ from cRPA directly without the empirical enhancement often employed to account for the frequency dependence. The value agrees with those used in \cite{Malcolms2026,Huang2017Stripes,Mai2024Fluctuating}, although sometimes larger $U$ values have been used \cite{barivsic2015high,PhysRevB.109.165111,85y6-2z4b}.
We further account for {a} double-counting correction and for the relative energy shift between $d$ and $p$ orbitals arising from neglected
$p$–$p$ interactions by an additional energy shift $\Delta_{\mathrm{dc}}\approx3.15\,\mathrm{eV}$
between $p$ and $d$ orbitals, following previous studies 
\cite{Weber2008,Kowalski2021}.
This $\Delta_{\mathrm{dc}}\approx3.15\,\mathrm{eV}$ needs to be subtracted from the Cu 3$d$ energy level $\varepsilon_d$  (or the one-particle charge transfer energy $\varepsilon_d-\varepsilon_p$  in Table~\ref{tab:cuprate_compounds}).
All calculations presented in  this paper are for the same $U = 6.5\,\mathrm{eV}$ and $\Delta_{\mathrm{dc}}\approx3.15\,\mathrm{eV}$.

To investigate superconductivity, we employ the dynamical vertex
approximation (D$\Gamma$A) in its $\lambda$-corrected ladder formulation \cite{Toschi2007,Katanin2009,RMPVertex}.
{To this end, we first calculate the irreducible vertex in the particle-hole channel using w2dynamics \cite{Wallerberger_CompPhysComm_2019_w2dynamics}. Next,}
we  use the DGApy code \cite{worm2023} with a momentum grid of \ej{at least $96\!\times\!96$ up to $130\!\times\!130$ for low hole dopings and temperatures}, and \ej{between 130 and 210} positive fermionic ($\nu,\nu'$) and an equal number bosonic Matsubara ($\omega$) frequencies \ej{$n_{\mathrm{core}}$} for the irreducible vertex $\Gamma_{ph}(\omega,\nu,\nu')$ in the particle-hole (ph) channel. The vertex is further supplemented by the bare $U$ in an outer box of \ej{430 to 510} Matsubara frequencies \ej{$n_{\mathrm{outer}}$}.
For additional details, see Ref.~\cite{Kitatani2022}.
To treat the multi-orbital Emery model {and superconductivity},
the D$\Gamma$A code needed to be extended, as described
in the Supplemental Material \cite{SM}.

\begin{table}[tb]
\centering
\caption{Different cuprate compounds with their corresponding conv.\ Emery model and downfolded Hubbard model hopping parameters.
}
\begin{tabular}{l|cccc|ccc}
\toprule
& \multicolumn{4}{c|}{conv.\ Emery model} 
& \multicolumn{3}{c}{Hubbard model} \\
\cmidrule(lr){2-5} \cmidrule(lr){6-8}
Material 
& $\epsilon_d - \epsilon_p$ & $t_{pd}$ & $t_{pp}$ & $t'_{pp}$ 
& $t$ & $t'/t$ & $t''/t$ \\
\midrule
Sr$_2$CuO$_2$Cl$_2$ \cite{Weber2012}
& 1.87 & 1.15 & 0.590 & 0.140 
& 0.41 & -0.070 & 0.124 \\
HgBa$_2$CuO$_4$ \cite{Weber2012}
& 1.93 & 1.25 & 0.649 & 0.161 
& 0.45 & -0.072 & 0.129 \\
Bi2212 \cite{Weber2012}
& 1.64 & 1.34 & 0.647 & 0.133 
& 0.49 & -0.056 & 0.131 \\
CaCuO$_2$ 
& 1.95 & 1.31 & 0.650 & 0.181 
& 0.46 & -0.065 & 0.132 \\
\bottomrule
\end{tabular}
\label{tab:cuprate_compounds}
\end{table}

\textit{Underestimation of effective hopping in the conventional Emery model.}
The  DFT-derived Wannier Hamiltonian $\bar{h}_{0}(\mathbf{k})$ 
is Fourier transformed to obtain the hopping elements between two lattice sites $\mathbf R$ and $\mathbf R'$, i.e., \lv{$\bar{h}_r(\mathbf{R-R}')$}.
In the conventional (conv.) Emery model only the three elements $t_{pd}$,  $t_{pp}$ and $t'_{pp}$ 
are considered. These are 
sketched in Fig.~\ref{fig:long_range_hoppings} and correspond to the dashed blue box.
While two of these, $t_{pd}$ and $t_{pp}$, are the largest hopping elements, the remaining ones are clearly not negligible.
In particular, they become relevant when considering longer-range hopping paths beyond nearest neighbors, which {in the conv.\ Emery model  requires}  combining several $t_{pd}$, $t_{pp}$, and \kh{$t'_{pp}$} hoppings.

{In Fig.~\ref{fig:long_range_hoppings}, we quantify the importance of the various hopping elements of the $dp$-model
through
$\Delta I= \frac{1}{n_{\mathbf k}}\sum_{\mathbf k}|\varepsilon({\mathbf k})-\mathbf{\hat{\varepsilon}(k)}|$,
i.e., the difference 
between (i) the full dispersion $\varepsilon({\mathbf k})$
(for the \ej{highest} energy band crossing the Fermi energy) that contains all hopping elements and (ii) $\hat{\varepsilon}({\mathbf k})$ where the respective hopping is explicitly excluded. The 1-norm is  normalized to the number of $\mathbf k$ points $n_{\mathbf k}$}.
\kh{The size of the orange circles in Fig.~\ref{fig:long_range_hoppings} clearly demonstrates that they are not negligible compared to $t_{pp}$ and $t'_{pp}$.} 

The insufficiency of the conv.\ Emery model \kh{with three hopping parameters only} also becomes apparent when downfolding it onto an effective one-band Hubbard model, as demonstrated in Table~\ref{tab:cuprate_compounds} for various cuprates. Clearly,
the next-nearest neighbor hopping $t'$ of the Hubbard model downfolded from the conv.\ Emery model is too small. \kh{Typical values for cuprates (and realistic one-band Hubbard model calculations) are rather in the range of 
$|t'/t|=0.15-0.25$}~\kh{\cite{Markiewicz2005,Qin2022,Kitatani2019}}.

\begin{figure}[tb!]
\centering
\includegraphics[width=1.\columnwidth]{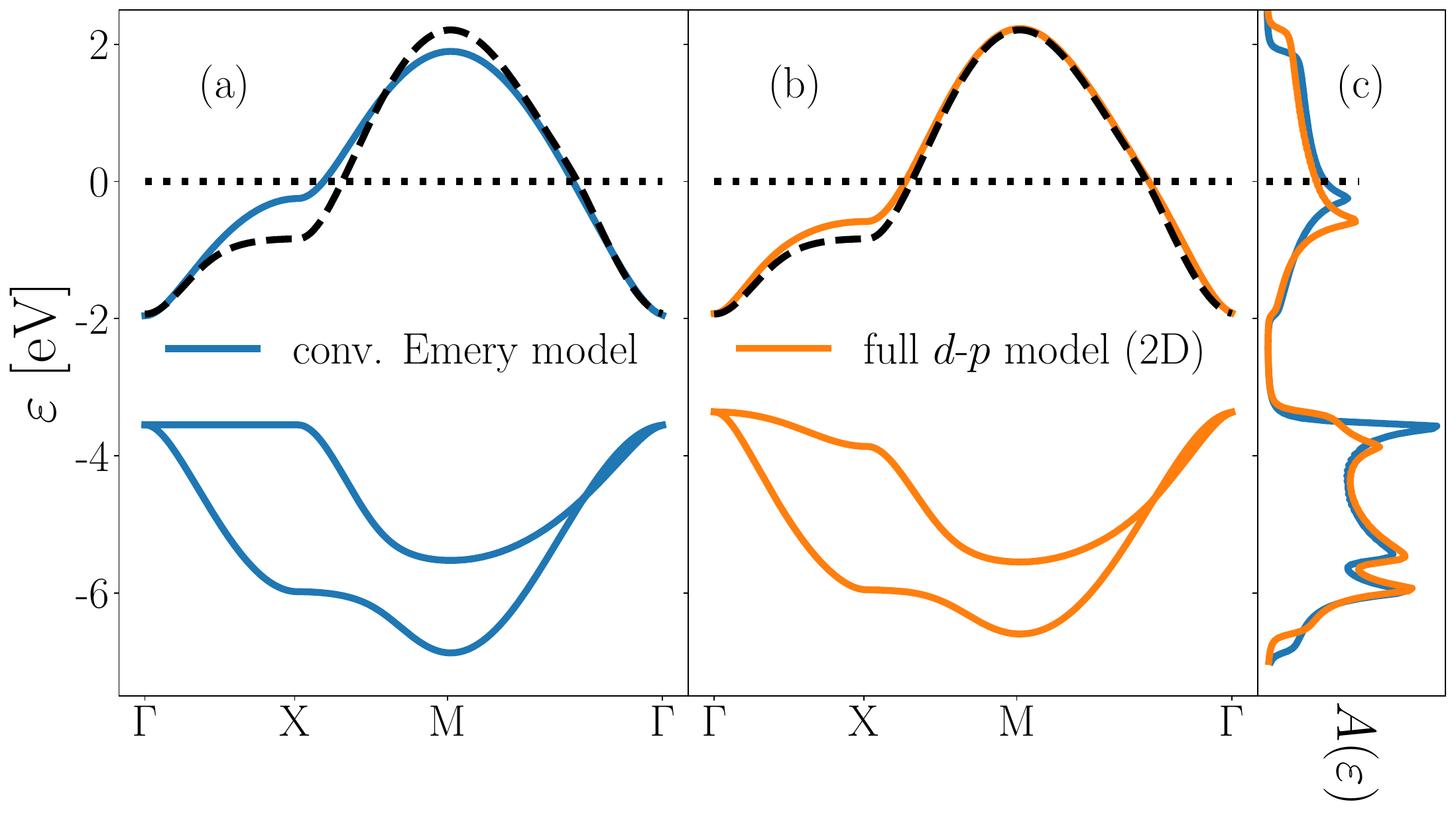}
\caption{Band structure of (a) the conv.\ Emery model and (b) the full $d$-$p$ model compared to the DFT-Wannier band (black dashed line). The corresponding density of states in panel (c) shows a clear shift of the van Hove singularity. }
\label{fig:Bandstructure_CaCuO2_comparison}
\end{figure}

This is also directly visible in the band structure \kh{in Fig.~\ref{fig:Bandstructure_CaCuO2_comparison} (a)}.
The longer range hoppings beyond the conv.\ Emery model and a larger $t'$ in the Hubbard model shift the van Hove singularity (inflection point at $X$ momentum, which for $t'=0$ is at the Fermi level) 
towards lower energies.
Clearly, for the conv.\ Emery model (blue line in Fig.~\ref{fig:Bandstructure_CaCuO2_comparison} (a)) the van Hove singularity is too close to the Fermi energy 
compared to the proper Wannier-(or DFT-)band (black dashed line).
This deficiency is overcome when including the hopping parameters beyond the conv.\ Emery model: orange dots and lines in Fig.~\ref{fig:long_range_hoppings} that lead to the orange band structure in Fig.~\ref{fig:Bandstructure_CaCuO2_comparison}(b)
\footnote{Note that the minor deviation of the full $d$-$p$ model (2D) 
from the DFT-Wannier bands originates from the hoppings in the $z$-direction which are neglected.}.

In Table~\ref{tab:CaCuO2_model_comparison} we  take a closer look at the particularly simple cuprate CaCuO$_2$. The conv.\ Emery model
gives a 3.5-times too small $t'$ whereas the full $d$-$p$ model agrees within acceptable 10\%, both in its 2D and its 3D version.

\begin{table}[tb]
\centering
  \caption{
Comparison between the downfolded one-band parameters of different three-band models (2D) for CaCuO$_2$ and the correct (direct) DFT Wannierization. The full $d$-$p$ model (2D) contains all in plane hoppings $\bar{h}_r(R_x,R_y,R_z=0)$\kh{;} the full $d$-$p$ model (3D) also all hoppings in the $z$ direction $\bar{h}_r(R_x,R_y,R_z)$.
}
\begin{tabular}{lccc}
\toprule
CaCuO$_2$ & $t$  & $t'/t$ & $t''/t$ \\
\midrule
DFT Wannier parameters & 0.49 & -0.25 & 0.20 \\
conv.\ Emery model & 0.46 & -0.07 & 0.13 \\
full $d$-$p$ model (2D) & 0.48 & -0.22 & 0.19 \\
full $d$-$p$ model (3D) & 0.48 & -0.27 & 0.20 \\
\bottomrule
\end{tabular}
\label{tab:CaCuO2_model_comparison}
\end{table}
Note that it is not possible to pinpoint exactly which hopping elements have to be included in addition to the conv.\ Emery model since many elements have a relevant contribution, cf.~Fig.~\ref{fig:long_range_hoppings}.  Let us also emphasize that fitting the conv.\ Emery model parameters to match the band structure does not work either\kh{, as there are only three hopping parameters
to fit  three bands  and already the band crossing the Fermi energy needs three hopping parameters $t$, $t'$ and $t''$ by itself}. 

From these considerations, we can conclude 
that 
relevant hopping elements are neglected when we restrict ourselves to the conv.\ Emery model.
While this conventional approach is appealing, and thus has been used extensively hitherto in the literature, it is not sufficient. In days of modern computing, it is also no problem to include all hopping elements of
the {\em full} $d$-$p$-model, which we make accessible in our data repository [link to be added upon publication].

\textit{Superconducting instabilities in cuprate $d$-$p$ models.}
We now \kh{come to the second main finding of our paper: the superconducting dome and its dependence on these longer-range hopping parameters}. 

Specifically, we compare the full $d$-$p$ model (2D)
to the conv.\ Emery model (both for CaCuO$_2$),  as well as to the so-called covalent Emery model. The latter is also a conv.\ Emery model. It was derived in \cite{Weber2012} for Bi$_2$Sr$_2$CaCu$_2$O$_8$ (Bi-2212)
with the hopping parameters in  \ej{Table~\ref{tab:cuprate_compounds}},
and is often employed in the relevant literature \cite{Kowalski2021}.
 For all of these models, 
we perform D$\Gamma$A calculations in the range of relevant hole dopings $\delta=5-n$ ($n$: total filling) for cuprate superconductivity.

\begin{figure}[tb!]
\centering
\includegraphics[width=1.\columnwidth]{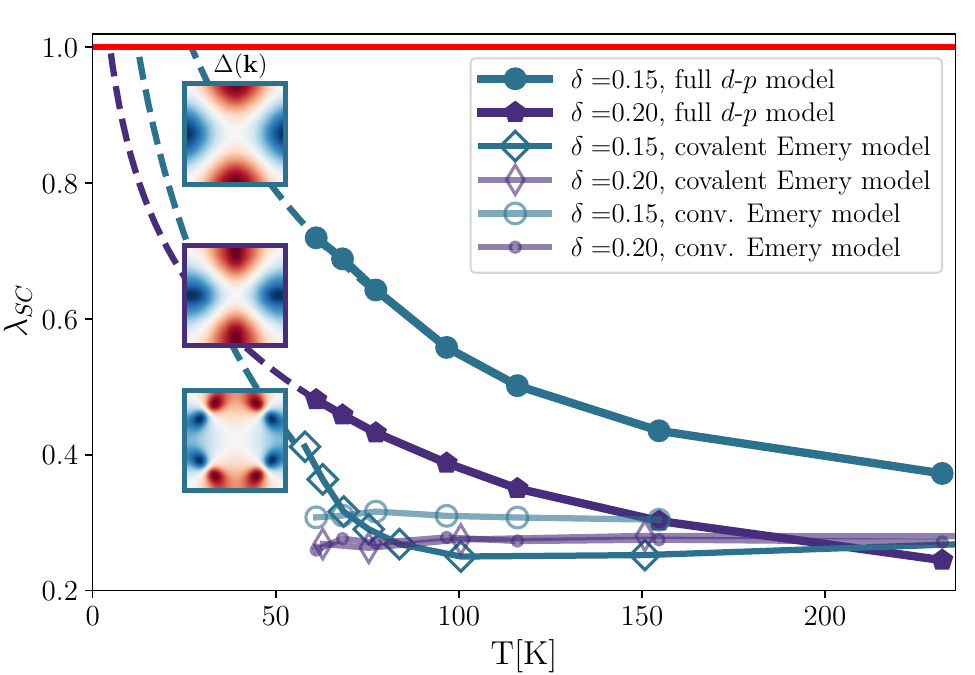}
\caption{Comparing the  superconducting $d$-wave eigenvalue $\lambda_{SC}(T)$ for the various models. The inset shows 
the symmetry of the 
superconducting eigenvector for the three cases showing a superconducting instability. 
}
\label{fig:lambda_Tremblay_vs_extended_emery}
\end{figure}
Figure~\ref{fig:lambda_Tremblay_vs_extended_emery} shows the
superconducting eigenvalue  $\lambda_{SC}(T)$,
with $\lambda_{SC}(T)=1$ indicating the superconducting instability; {additional dopings are shown in the Supplemental Material \cite{SM}.}
The conv.\ Emery model (empty circles) shows no sign  {yet}
of an increase of   $\lambda_{SC}(T)$ down to the lowest accessible temperatures \kh{and for a doping of 15\% and 20\%, i.e., the typical range of optimal doping in cuprates}.
This changes dramatically if we include the hopping elements beyond the conv.\ Emery model (\ej{two uppermost} solid lines), where we obtain superconductivity at $\delta=0.15$. At  $\delta=0.2$,
we also extrapolate to a small $T_c$, using the fit formula  \ej{$\lambda_{SC}(T)=a+b\log{T}$} as in previous calculations \cite{Kitatani2020,Kitatani2022b,yd8w-frs8,wu2026single}.
Although for this latter doping, also $T_c=0$ is possible within the extrapolation error.
In any case, we clearly see how important the longer range hopping 
beyond the conv.\ Emery model is.

Also shown in Fig.~\ref{fig:lambda_Tremblay_vs_extended_emery}
is  $\lambda_{SC}(T)$ for the covalent Emery model for Bi-2212.
Here, $\delta=0.2$ behaves similar as for the conv.\ Emery model (for CaCuO$_2$). For both,   $d$-wave superconductivity is the leading eigenvector, but there is no indication at all of a
superconducting transition. 

For the covalent Emery model at  $\delta=0.15$ we
see a rather sudden and sharp increase of the $d$-wave eigenvalue at low temperatures. {The associated $d$-wave eigenfunction, shown in the inset of   Fig.~\ref{fig:lambda_Tremblay_vs_extended_emery},
shows a somewhat unusual modulation with a suppression at the antinodes.}
This can be explained as follows: without longer range hoppings, the van Hove singularity is too close to the Fermi energy ($t'$ correspondingly small in Table~\ref{tab:cuprate_compounds}) so that a pseudogap
already starts to develop as discussed previously in \cite{Malcolms2026}.  
This pseudogap suppresses the spectral weight at the antinode 
which, thus, contributes little to superconductivity.
In the inset of Fig.~\ref{fig:lambda_Tremblay_vs_extended_emery}
we see that the superconducting pairing adjusts to this suppressed weight at the antinode by developing a modulated $d$-wave order parameter with a suppressed superconducting eigenvector at the antinode where the spectral function is suppressed
\footnote{
This effect is possibly somewhat exaggerated in ladder D$\Gamma$A,
since the (three-leg) vertex can become enhanced ($\mathbf k$-dependent) promoting superconductivity further into the pseudogap regime, see 
\cite{Yu2025} where this effect was observed, albeit at a considerably smaller doping. Let us also note, that 
 we observe a competing $p$-wave eigenvalue for this parameter set. 
}.

\begin{figure}[tb!]
\centering
\includegraphics[width=1.\columnwidth]{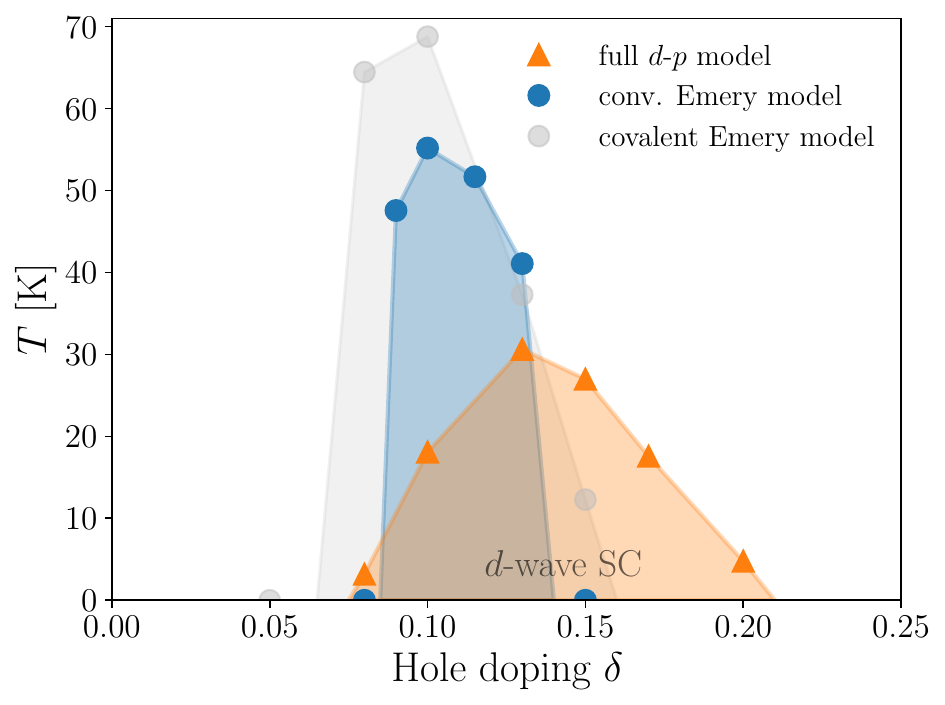}
\caption{Phase diagram as a function of temperature and hole doping showing the $d$-wave superconducting dome of the full $d$-$p$ model, the conv. and covalent Emery model.}
\label{fig:D_wave_dome}
\end{figure}

In Fig.~\ref{fig:D_wave_dome} we present the
superconducting dome of the full $d$-$p$ model (2D) for CaCuO$_2$.
We observe the typical dome of cuprate superconductors with superconductivity in the range of
about 7-22\% hole doping. \kh{Let us note, however, that e}xperimentally bulk CaCuO$_2$ has hitherto not been found to be superconducting \cite{raveau1997mechanisms}\kh{.
But,} superconducting heterostructures could be synthesized \cite{PhysRevLett.115.147001,PhysRevLett.92.157006}. In films,  CaCuO$_2$ has a maximal $T_c$ of about 40\,K \cite{DiCastro2014,PhysRevLett.115.147001}. 

Let us emphasize that our calculations are for bulk CaCuO$_2$, and differences between films and bulk which are hitherto poorly understood might be of importance.
\kh{In contrast, the conv.\ Emery model (blue curve in
Fig.~\ref{fig:D_wave_dome}) has a very narrow superconducting dome between 9\% and 14\% hole doping only. This demonstrates the importance of long-range hopping parameters for the quantitative description of superconductivity in cuprates with the Emery model.}


 Fig.~\ref{fig:D_wave_dome} also shows the  superconducting dome for the covalent Emery model, which is
 the conv.\ Emery model for Bi2212. The
 main difference to the conventional Emery model for CaCuO$_2$ is the larger charge transfer energy 
 $\varepsilon_d-\varepsilon_p$, see Table~\ref{tab:cuprate_compounds}. 
 In Fig.~\ref{fig:lambda_Tremblay_vs_extended_emery}, we have seen that an atypical, i.e.,  modulated, $d$-wave order parameter 
 is found for this model (but not in experiment \cite{Vishik2010})
because the van Hove singularity is  too close to the Fermi level
\kh{(which leads to a large pseudogap, eventually suppressing plain $d$-wave superconductivity)}. \kh{The same modulation is also found for the conventional Emery model (blue line in Fig.~\ref{fig:D_wave_dome}).} 
\kh{That is, without longer range hoppings,  the  van Hove singularity is too close to the Fermi level. This again
leads to a modulated $d$-wave order parameter and} a rather narrow superconducting dome shifted to smaller hole-dopings. \kh{In particular,} there is no superconductivity at larger dopings above 17\% for the covalent Emery model for Bi2212, whereas in experiment superconductivity for Bi2212 extends up to 30\%
 hole doping~\cite{Drozdov2018}.

\textit{Conclusions.}
Superconductivity in the conventional Emery model has hitherto only been studied for 
\kh{small} clusters that are known to largely overestimate $T_c$---at least  for the Hubbard model. 
We here calculate superconductivity of the full $d$-$p$ model 
using D$\Gamma$A, and find a phase diagram in
good agreement with cuprates\kh{. F}or an accurate quantitative description,
hopping matrix elements beyond the $t_{pd}$, $t_{pp}$ and \kh{$t'_{pp}$} of the conventional Emery model are absolutely necessary. \kh{Otherwise, the superconducting dome is found at too small hole doping and there is also a strong} modulation of the $d$-wave order parameter.
Our work opens the path to a more material realistic description of cuprate superconductivity which certainly necessitates the inclusion of oxygen orbitals.

\begin{acknowledgements}
\emph{Acknowledgments---}We acknowledge very helpful discussion with
Simone di Cataldo, Motoharu Kitatani, and Juraj Krsnik.
We further acknowledge funding through the Austrian Science Funds (FWF) Project Grant DOI 10.55776/I5398 and the European Research Council through ERC-2024-ADG RealSuper project DOI 10.3030/101201037, from the Deutsche Forschungsgemeinschaft (DFG) through research unit
FOR5249 ``QUAST'' project No.~449872909 (Projects P1 and P4), partially funded by the FWF-funded as (sub)project DOI 10.55776/KIN2563725,
and from the FWF Spezialforschungsbereich (SFB) QM\&S 
project DOI 10.55776/F86.
L.~S.~acknowledges support from the National Natural Science Foundation of China (Grant No.~12422407).
Calculations have been done on the Vienna Scientific Cluster (VSC).

This project is funded in part by the European Union. Views and opinions expressed are however those of the author(s) only and do not necessarily reflect those of the European Union or the European Research Council Executive Agency. Neither the European Union nor the granting authority can be held responsible for them.

For the purpose of open access, the authors have applied a CC BY public copyright license to any Author Accepted Manuscript version arising from this submission.

\emph{Data availability---}The data that support the findings of this article are openly available at [link to be provided with publication].
\end{acknowledgements}

\bibliography{main.bib,main_2.bib}

\end{document}


\title{\textit{Supplemental Material:} \kh{\it ``Beyond the conventional Emery model: \\ crucial role of long-range hopping for cuprate superconductivity"}
}

\author{Eric Jacob}
\email{eric.jacob@tuwien.ac.at}
\affiliation{Institute of Solid State Physics, TU Wien, 1040 Vienna, Austria}

\author{M.~O.~Malcolms}
\affiliation{Max-Planck-Institut für Festkörperforschung, Heisenbergstraße 1, 70569 Stuttgart, Germany}

\author{Viktor Christiansson}
\affiliation{Institute of Solid State Physics, TU Wien, 1040 Vienna, Austria}

\author{Leonard M. Verhoff}
\affiliation{Institute of Solid State Physics, TU Wien, 1040 Vienna, Austria}

\author{Paul Worm}
\affiliation{Institute of Solid State Physics, TU Wien, 1040 Vienna, Austria}

\author{Liang Si}
\affiliation{School of Physics, Northwest University, Xi'an 710127, China}
\affiliation{Institute of Solid State Physics, TU Wien, 1040 Vienna, Austria}

\author{Philipp Hansmann}
\affiliation{Department of Physics, Friedrich-Alexander-Universit\"at Erlangen/N\"urnberg, 91058 Erlangen, Germany}
\affiliation{Science Institute and Faculty of Physical Sciences, University of Iceland, 107 Reykjav\'ik, Iceland}

\author{Thomas~Sch{\"a}fer}
\affiliation{Max-Planck-Institut für Festkörperforschung, Heisenbergstraße 1, 70569 Stuttgart, Germany}
\affiliation{Dipartimento di Fisica, Università di Trieste, Strada Costiera 11, I-34151 Trieste, Italy}

\author{Karsten Held}
\email{held@ifp.tuwien.ac.at}
\affiliation{Institute of Solid State Physics, TU Wien, 1040 Vienna, Austria}

\begin{abstract}
    In this supplemental material we provide additional information. Specifically, we show the density functional theory band structure in Section~\ref{Ssec:DFT}; and discuss the necessary modifications to calculate superconductivity in $d$-$p$ models in Section~\ref{SSec:calc}. Further, in Section~\ref{SSec:lambda} we plot the superconducting eigenvalues for all dopings and models studied; in Section~\ref{SSec:PG} we present the pseudogap near the expected optimal doping of cuprates for the covalent Emery model; and in Section~\ref{SSec:FS} the Fermi surfaces for the full $d$-$p$ model.
\end{abstract}
\maketitle

\section{Band Structure and Wannier projection}
\label{Ssec:DFT}
\begin{figure}[tb]
\centering
\includegraphics[width=.8\columnwidth]{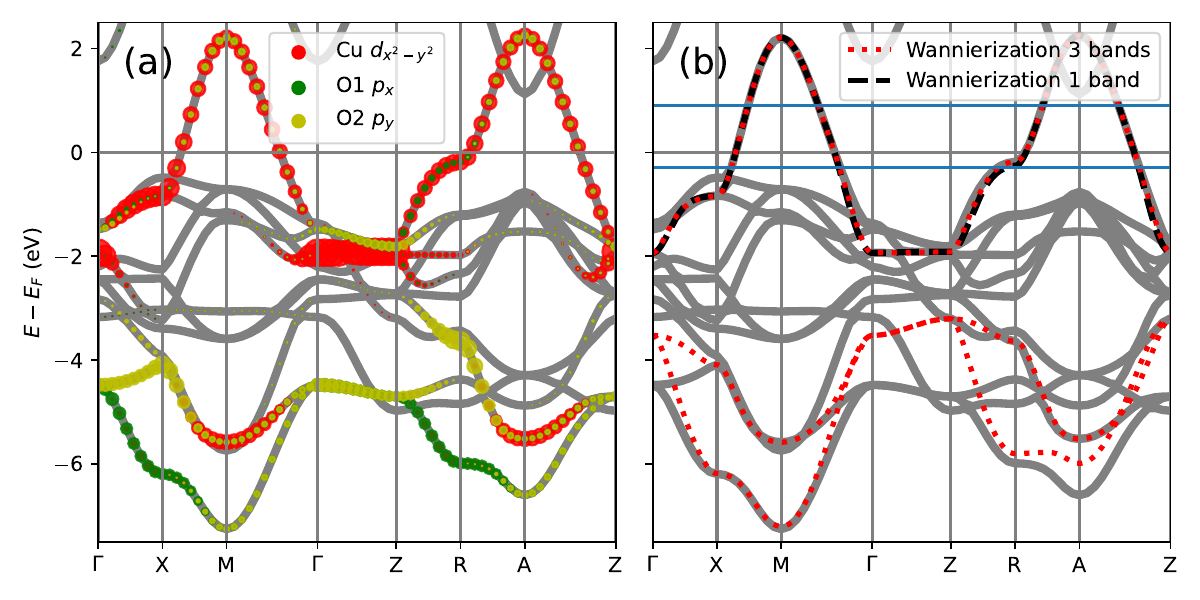}
\caption{DFT bands of CaCuO$_2$ (gray). Superimposed are: (a) Projections onto Cu $d_{x^2-y^2}$ and the in-plane $p_{x}$ and $p_y$ orbitals pointing towards it (cf.\ main text Fig.~1). 
(b) Wannier Hamiltonian of 3 band (red) and 1 band (black) model. Blue, horizontal lines indicate the \textit{frozen window} used within \textsc{Wannier90} for both models.
}
\label{fig:dft CaCuO2}
\end{figure}
The density functional theory (DFT) band structure of CaCuO$_2$ (see main text for computational details) along a high-symmetry path through the full 3D Brillouin zone is shown in both panels of Fig.~\ref{fig:dft CaCuO2}. 
The left panel shows the projections of DFT bands onto the three local atomic orbitals that are relevant to the \ej{$d$-$p$ models considered in the main text}. The single band crossing Fermi level shows a large Cu $d_{x^2-y^2}$ character (red). However, it is also strongly hybridized with lower-energy O $p_x$ (green) and $p_y$ states (yellow) of the oxygen atoms neighboring Cu in the $x$- and $y$-direction, respectively. The lobes of these two $p$ orbitals point towards Cu and therefore overlap with the Cu $d_{x^2-y^2}$ orbital.
The right panel in Fig.~\ref{fig:dft CaCuO2} compares the Wannier Hamiltonian of the 1 band model and the 3 band model to the DFT Bloch bands. The 1 band model is sufficient to successfully reproduce the band crossing Fermi level. However, the strong hybridizations between $d$ and $p$ states, as well as the large spread of the orbital in the 1 band model, suggest that an elaborated and material realistic model for cuprates needs more than a single localized Cu $d_{x^2-y^2}$ orbital.

\ej{In the main text we perform superconducting calculations using models based on DFT for effectively two-dimensional systems, with momentum grids of the form $n_{k_x}\times n_{k_y}\times 1$. When transforming the real-space Wannier Hamiltonian to momentum space, matrix elements with an out-of-plane component may contribute. We therefore distinguish between two DFT-based models for CaCuO$_2$: the full $d$-$p$ model (3D) and the full $d$-$p$ model (2D). The former is obtained by directly transforming the full Wannier Hamiltonian, whereas for the latter all matrix elements with an out-of-plane component are set to zero prior to the transformation. Except for Table~II of the main text, we use the full $d$-$p$ model (2D) and thus omit this distinction. For the conventional Emery model, no such distinction arises by construction, since it contains no out-of-plane hoppings.
}

\section{Superconducting calculations for $d$-$p$ models}
\label{SSec:calc}
The details of the ladder D$\Gamma$A extension to treat $d$-$p$ models are described in the Supplemental Information of \cite{Malcolms2026} up to the point where $\Sigma_{\text{D}\Gamma\text{A},dd}^{\text{ladder}}\left(\mathbf{k},i\nu\right)$ is calculated. To determine the superconducting $T_c$, we have to go beyond this step and analyze the pairing instability. The pairing susceptibility has an orbital structure that originates only from the external non-interacting legs. The reason is that the irreducible particle-particle vertex $\Gamma_{\text{pp},ijkm}$, where $i,j,k,m$ denote orbital indices, has only one non-vanishing contribution, $\Gamma_{\text{pp}}\equiv\Gamma_{\text{pp},dddd}$. The reason for this is that all Feynman diagrams for $\Gamma$ must start at an interaction $U$, which is restricted to the only interacting Cu 3$d_{x^2-y^2}$ orbital. Since the superconducting instability cannot originate from the external Green function ''legs'', we can analyze the linearized Eliashberg equation in a one-orbital form
\begin{equation}
\label{eq: Eliashberg}
\lambda_{SC}\Delta(k)=-\frac{1}{2\beta}\sum_{k'} \Gamma_{\text{pp}}^{q=0}(k,k')  G_{dd}(k') G_{dd}(-k')\Delta(k'),
\end{equation}
where $\lambda_{SC}\to 1$ determines $T_c$. The Green’s function $G_{dd}$ is obtained by inserting $\Sigma_{\text{D}\Gamma\text{A},dd}^{\text{ladder}}\left(\mathbf{k},i\nu\right)$ into the three-orbital Dyson equation, as already described in \cite{Malcolms2026}, and taking the $dd$ orbital component. The vertex $\Gamma_{\text{pp}}^{q=0}(k,k')$ is computed in complete analogy to the Hubbard model, as detailed in the Supplemental Material of \cite{Kitatani2019}, by taking the $d$ components throughout. As for the Hubbard model, the explicitly given formulas either have to be (anti)symmetrized in $k'$ for the singlet (triplet) channel, or a solver that only allows solutions $\Delta(k)$ with the correct symmetry has to be used to recover both contributions from the $ph$ and $\bar{ph}$ channels.

\section{Doping dependence $\lambda_{SC}$ for the full $d$-$p$, covalent Emery model and conventional Emery model}
\label{SSec:lambda}

To determine the $d$-wave superconducting dome for the full $d$-$p$ model, the covalent and conventional Emery model, we performed calculations for additional fillings beyond those explicitly shown in the main text. We present these results here and further note that the gap functions of the covalent Emery model exhibit an even stronger suppression (modulation of the $d$-wave)  at the antinodal point for dopings $\delta=0.13, 0.10, 0.08$ at the temperature (155 K) for which we show the gap function at $\delta=0.15$ in the main text. This trend is consistent with the stronger pseudogap at these fillings in the covalent model. The conventional Emery model shows a similar behavior. For the full $d$-$p$ model, the gap functions remain similar across all fillings and do not show an additional filling dependence beyond that already illustrated in the main text.

\begin{figure}[h!]
\centering
\includegraphics[width=0.45\columnwidth]{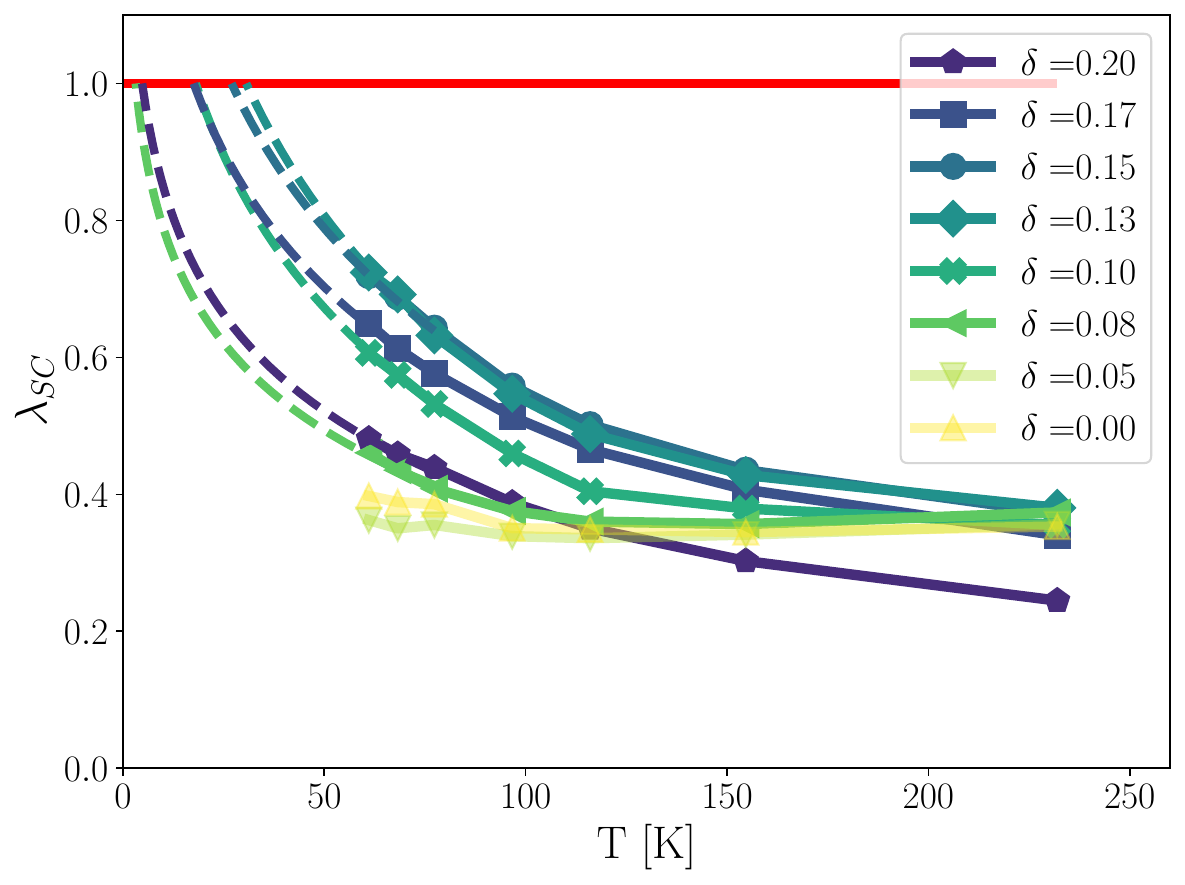}
\caption{Largest d-wave superconducting eigenvalues for the full $d$-$p$ model for all calculated values of the doping $\delta$. The dashed lines show the extrapolation to obtain $T_c$. }
\label{fig:lambda_dopings}
\end{figure}

\begin{figure}[h!]
\centering
\includegraphics[width=0.45\columnwidth]{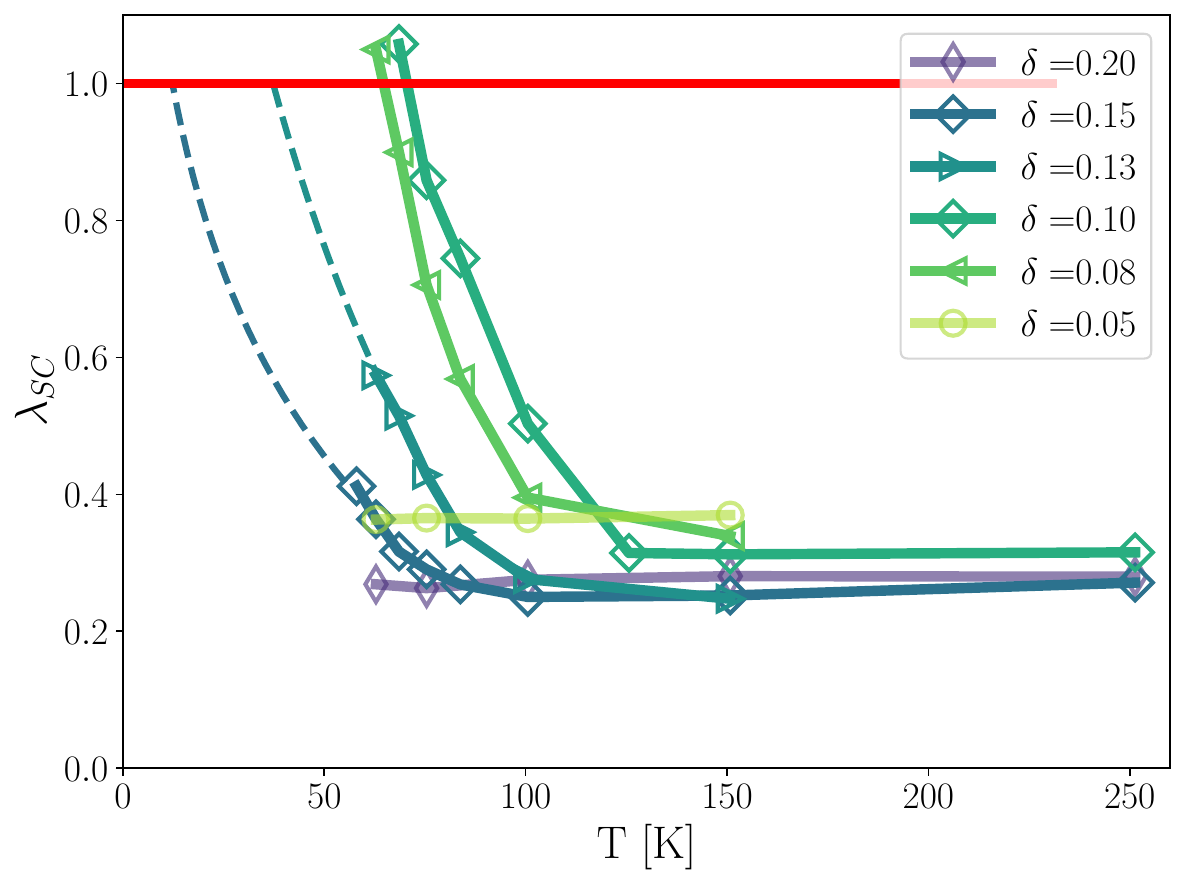}
\caption{Same as \protect Fig.~\ref{fig:lambda_dopings} but for the covalent Emery model. }
\label{fig:lambda_dopings_covalent}
\end{figure}

\begin{figure}[h!]
\centering
\includegraphics[width=0.45\columnwidth]{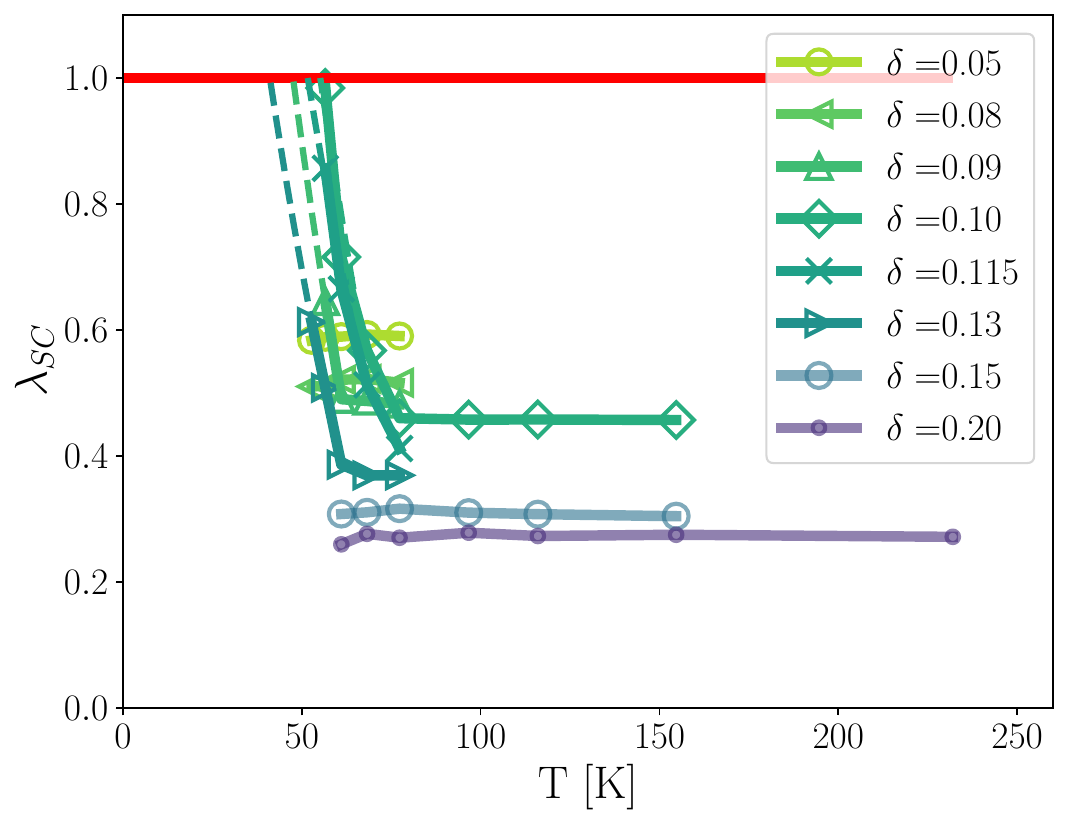}
\caption{Same as \protect Fig.~\ref{fig:lambda_dopings} but for the conv. Emery model.  }
\label{fig:lambda_dopings_conventional}
\end{figure}
\clearpage
\pagebreak

\section{Pseudogap in the covalent Emery model for $\delta=0.15$ at lower temperatures}
\label{SSec:PG}

\begin{figure}[h!]
\centering
\includegraphics[width=0.6\columnwidth]{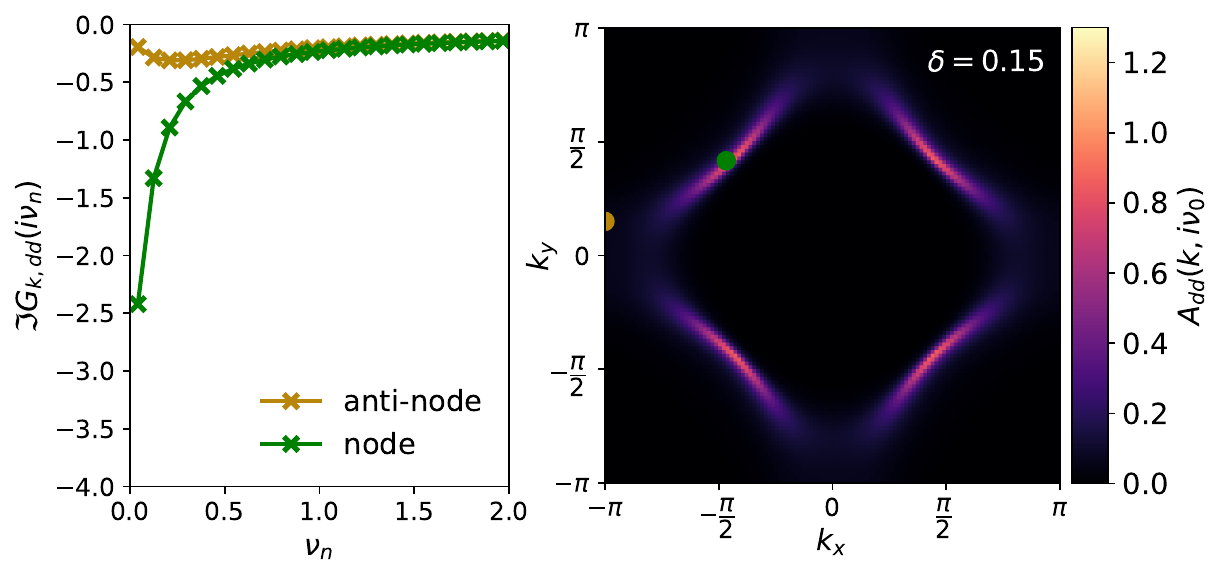}
\caption{Imaginary part of the $d$-$d$ Green's function for two selected $\mathbf{k}$ points (left) and  $\mathbf{k}$-resolved spectral-function 
at the Fermi energy (the momenta from the left panel are marked in the same color). Note, we take the spectral function 
$A({\mathbf k},i\nu_0)$ at the lowest Matsubara frequency $\nu_0=\pi/T $ ($T=\frac{1}{75}t_{pp}^{-1}\approx100\,\mathrm{K}$) to avoid analytic continuation.}
\label{fig:ImG_node_antinode}
\end{figure}

In the main text we attribute the additional suppression of the gap functions for $\delta=0.15$ at the anti-nodal point to the suppression of the spectral weight at this point in the pseudogap phase. In \cite{Malcolms2026} $\delta=0.15$ lies on the boundary of the shaded pseudogap phase for the explored temperatures. For the, in comparison, much lower temperatures explored \kh{here}, we also observe for $\delta=0.15$ a strong nodal–antinodal dichotomy. 
This can be seen in Fig.~\ref{fig:ImG_node_antinode} where we only plot the orbital $d$-$d$ components, since this is the quantity directly entering our superconducting analysis.

\section{Fermi surfaces in the superconducting range for the full $d$-$p$ model}
\label{SSec:FS}

In the range of dopings $\delta=0.08$--$0.20$, unlike for the covalent Emery model, we do not observe a modulation of the $d$-wave gap functions in the full $d$-$p$ model. The Fermi surfaces in this range are also different and show a much weaker suppression of the spectral weight at the antinode. In addition to the comments in the main text, we explicitly show Fermi-surface proxies that are directly relevant to our superconducting calculations.

\begin{figure}[h!]
\centering
\includegraphics[width=.85\columnwidth]{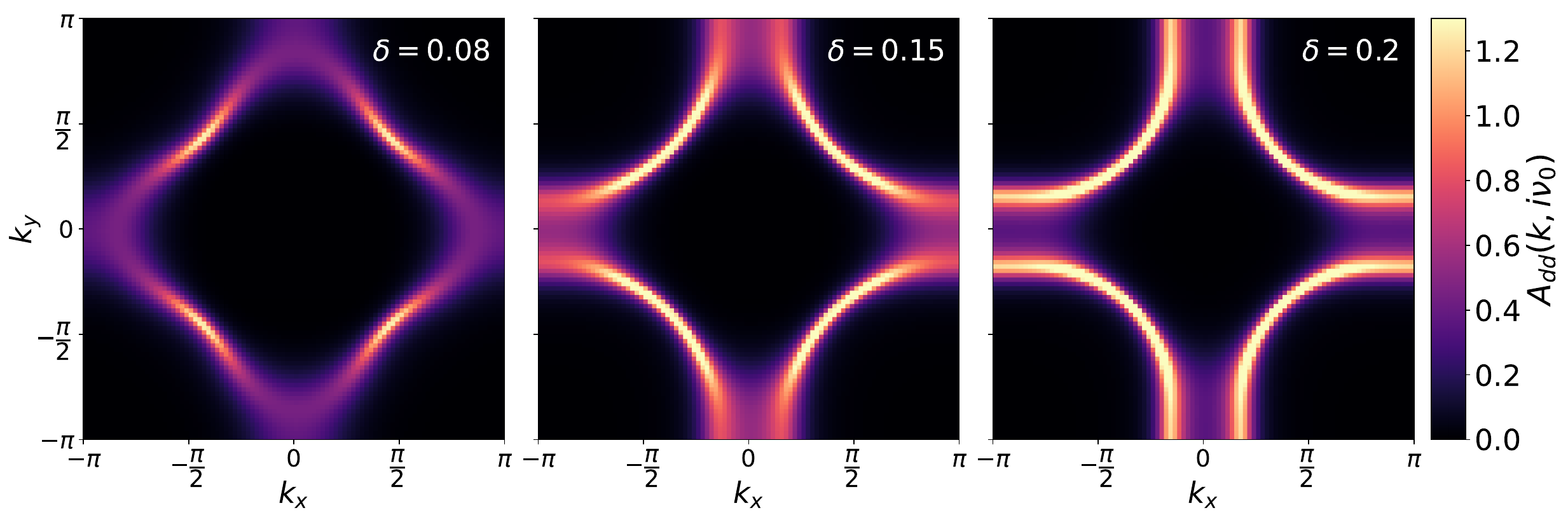}
\caption{$\mathbf{k}$-resolved spectral-function proxy for the full $d$-$p$ model for three different dopings within the superconducting range at $T=\frac{1}{120}\,\mathrm{eV}^{-1}\approx97\,\mathrm{K}$.}
\label{fig:FS_full_dp}
\end{figure}
\FloatBarrier
\bibliography{main.bib,main_2.bib}